# Alignment behaviour of 2D Diopsides (d-silicates) under the influence of acoustic strain

Himakshi Mishra[a], Surbhi Slathia[b], Bruno Ipaves[c], Raphael Benjamim de Oliveira[c], Marcelo Lopes Pereira Júnior[d,e], Raphael Matozo Tromer[f], Gelu Costin[g], Nicholas R. Glavin[h], Ajit K. Roy[h], Douglas Soares Galvão[c,*], and Chandra Shekar Tiwary [a,*]

[a] *Metallurgical and Materials Engineering, Indian Institute of Technology Kharagpur, Kharagpur 721302, India.*

[b]*School of Nano Science and Technology,*

*Indian Institute of Technology Kharagpur, Kharagpur 721302, India*

[c]*Applied Physics Department and Center for Computational Engineering & Sciences, State University of Campinas, Campinas, SP, 13083-970, Brazil.*

[d]*College of Technology, University of Brasília, Brasilia, DF, 70910-900, Brazil.*

[e]*Materials Science and Nano Engineering Department, Rice University, Houston, TX, 77005, USA.*

[f]*Institute of Physics, University of Brasília, Brasilia, DF, 70910-900, Brazil.*

[g]*Department of Earth, Environmental and Planetary Sciences, Rice University, Houston, TX 77005, United States*

[h]*Materials and Manufacturing Directorate, Air Force Research Laboratory, Wright-Patterson Air Force Base, Dayton, OH, United State*

* *galvao@ifi.unicamp.br*

**chandra.tiwary@metal.iitkgp.ac.in*

**Abstract:**

Controlling the alignment of two-dimensional (2D) materials is crucial for optimizing their electronic and mechanical properties in next-generation devices. This study explores how electric fields can manipulate the orientation of 2D diopside ($CaMgSi_2O_6$) flakes—a flexible silicate material—through a phenomenon called flexoelectricity, where applied voltage generates mechanical strain. We exfoliated diopside crystals into ultrathin flakes, placed them on microelectrodes, and used AC electric fields to induce alignment via acoustic strain. Raman spectroscopy showed that the flakes reoriented/realigned under the field, with vibrational peaks weakening most at high frequencies (10 MHz). Electrical tests revealed this alignment improves conductivity by 20-30%, as straightened flakes create better pathways for current flow. Molecular dynamics simulations further explained how these flakes naturally align on surfaces within picoseconds, matching our experimental observations. Together, these findings demonstrate a practical way to tune diopside properties using electric fields, opening doors for its use in flexible electronics, sensors, and energy devices.

## 1. Introduction:

The field of two-dimensional (2D) materials has experienced a surge in research, primarily driven by their potential in miniaturized electronic, optical, and mechanical applications [1, 2, 3]. The combination of easy fabrication, unique physical properties, and new phenomena arising from 2D electronic confinement has attracted significant attention in materials science. Variations in atomic arrangement within these materials play a crucial role, as such features can either enhance or degrade their properties depending on the application [4]. The discovery of 2D silicates began with the exfoliation of naturally occurring layered materials like montmorillonite and kaolinite, known for their weak interlayer bonding [5,6] and swelling [7] properties. While early research focused on van der Waals solids [8], advancements revealed that non-layered materials could also produce 2D structures [9]. Representing ~95% of the Earth's crust, silicates are abundant and structurally diverse, encompassing layered and non-layered structures. Recent methods such as hydrogenation, Al-doping, and Molecular Beam Epitaxy (MBE) fabrication have facilitated the exfoliation of non-layered silicates, like diopsides ($CaMgSi_2O_6$), a monoclinic pyroxene with exceptional mechanical strength and biocompatibility.

Works by Novoselov et al. and others [10, 11] have demonstrated how the electronic properties of these materials could be tuned by external factors, such as strain, making them excellent candidates for flexible electronics. As interest expanded, 2D materials such as diopsides have unveiled their remarkable potential, exhibiting exceptional mechanical flexibility and unique electronic properties [12]. These materials present new perspectives in expanding the family of 2D systems beyond traditional candidates like graphene, offering new ways for exploiting applications in flexible electronics, optoelectronics, and energy storage.

Notably, alignment/stacking of 2D materials plays a critical role in addressing the unique challenges that arise when these materials are integrated into nanoscale devices, particularly as scaling in traditional materials becomes increasingly challenging. Alignment helps prevent inconsistencies in charge transport and thermal management that become more pronounced at smaller scales, where random orientation could lead to performance variability and inefficiencies. In addition, uniform alignment across large-scale fabrication allows for reliable device functionality, which is crucial for scalable production. Although 2D materials naturally overcome some scaling limitations due to their thinness and reduced dimensionality, precise alignment is essential for ensuring consistent performance in miniaturized applications. Properly aligned 2D layers improve electron mobility, thermal conductivity, and mechanical flexibility, which are crucial as device dimensions shrink.

In 2017, Lin et al. [13] demonstrated a novel approach for aligning graphene sheets using a rotating magnetic field. Their method allowed for controlled, planar alignment of the graphene layers, enhancing the material's properties for electronic and optoelectronic applications. This approach enabled more efficient use of graphene's remarkable characteristics, such as high conductivity and mechanical strength. The findings underscored the potential of magnetic field-assisted alignment techniques in advancing the functionality and scalability of graphene in various 2D material applications. In 2007, Lim et al. [14] developed a solution-based acoustic manipulation technique utilizing ultrasound-generated forces and streaming to achieve the parallel alignment of carbon nanotubes (CNTs) on solid substrates. Their study, supported by a theoretical model analyzing forces and torques, specifically focused on CNTs but also highlighted principles and methods that could potentially be adapted for manipulating other 2D materials.

With the exploration of 2D diopsides, their research has increasingly focused on the flexoelectric effect and the interplay between electric fields and strain generation. This phenomenon offers promising opportunities for applications such as energy harvesting [15] and wearable and health monitoring devices [16,1]. The flexoelectric properties of 2D materials enable them to induce acoustic strain within their structure when subjected to an electric field. This self-induced strain arises from the coupling between the electric field and the material's inherent flexoelectric response, offering unique opportunities for manipulating their physical properties without external mechanical input. This interaction of electric fields with acoustic strain and mechanical deformation caused by vibrations can control alignment within 2D layers. This method could precisely align the material at the atomic scale, a critical factor in achieving optimized performance for specific applications.

The present study investigates the alignment behavior of 2D diopsides (d-silicates) under the influence of an AC electric field, which generates acoustic strain through flexoelectric effects. The unique lattice structure and mechanical flexibility of d-silicates provide an ideal platform to explore how flexoelectricity influences material alignment. The study aims to deepen our understanding of alignment mechanisms in 2D systems and their potential implications for advanced electronic applications by focusing on these properties. We have also used fully atomistic molecular dynamics (MD) simulations to gain further insights into the atomic structure and electronic behavior. This versatile approach can be effectively applied to examine different silicate-based materials.

## 2. Methods:

*2.1 Experimental Details:*

Bulk diopside crystals ($CaMgSi_2O_6$) were obtained from the Kimberley region of South Africa. The mineral exhibits a characteristic greenish color. While $CaMgSi_2O_6$ represents the ideal stoichiometric composition of diopside, the exfoliated diopside used in this study was derived from the natural bulk sample from Kimberley, whose elemental composition is provided in Table SX1 of the supplementary information. These crystals were initially ground into a fine powder using a mortar and pestle. 50 mg of the powdered material was dispersed in 50 mL of isopropyl alcohol (IPA) and subjected to probe sonication for 5 hours. To ensure the temperature of the solvent remained below 35°C during the sonication process, the operation was paused every 30 minutes for 15–20 minutes, allowing the solution to cool. After sonication, a stable suspension containing exfoliated 2D diopsides was collected and preserved for subsequent experiments.

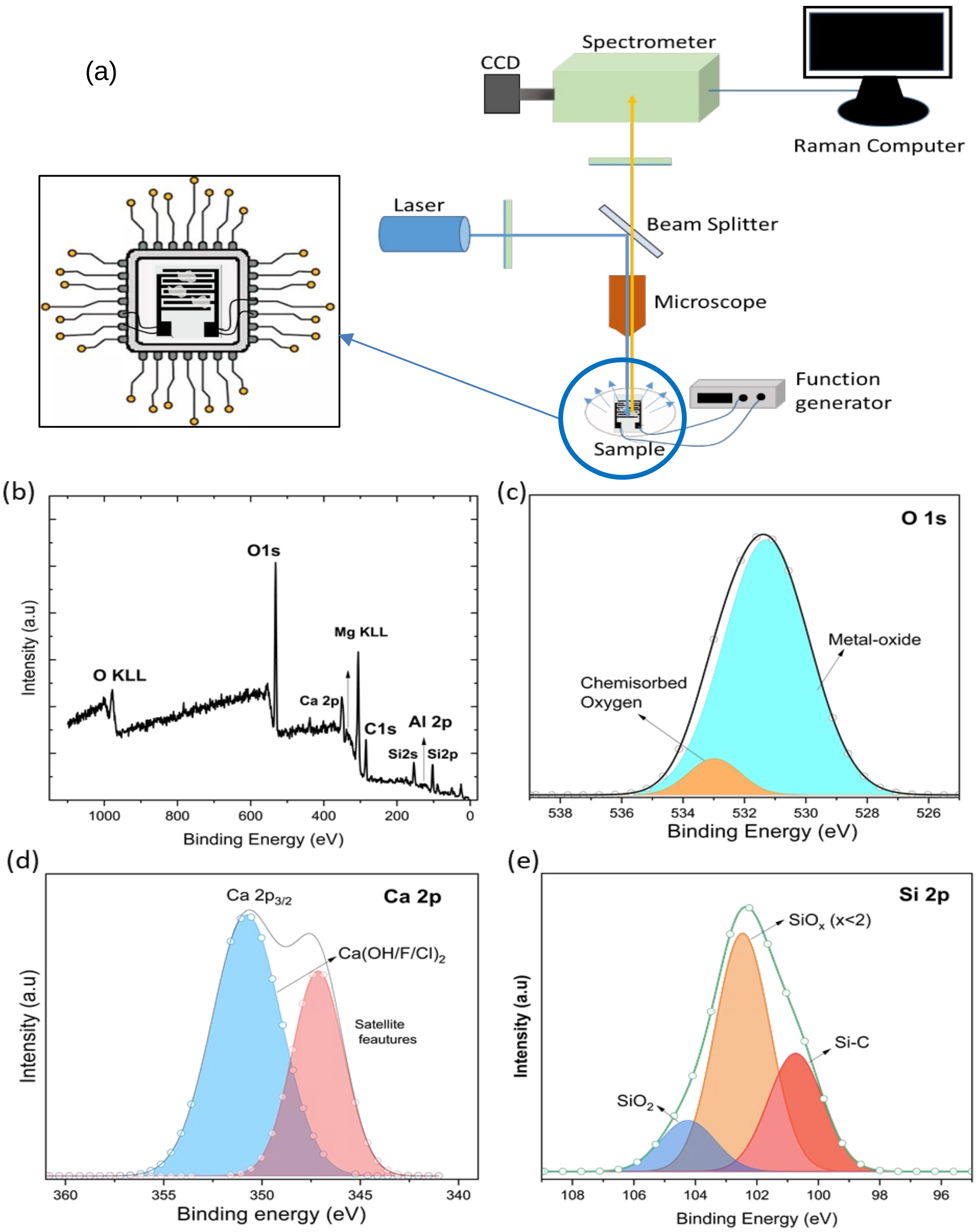

**Figure 1**. (a) Schematic of the optical setup for Raman characterisation of the interdigitaded electrodes (IDE) mounted over an IC chip coated with 2D diopsides ($CaMgSi_2O_6$) connected to an AC signal, (b-c) XPS measurements confirming the elemental composition and chemical states of 2D diopsides.

For device fabrication, thin-film gold IDEs on an alumina ceramic thin film sheet with a finger gap and width of 50 µm were used. A small amount of the 2D diopside solution was drop-cast onto the IDE sheet and heated on a hot plate set to 100°C until the solvent evaporated, leaving a uniform material coating. This process was repeated multiple times to ensure that a continuous layer of the exfoliated diopsides formed a bridge between adjacent electrodes. The device was then mounted over an IC chip (**Figure 1**) and connected to a digital power supply (Tektronix) to apply an alternating current (AC) electric field. This arrangement induces acoustic strain in the material via the electric field for the alignment experiments.

The study utilized both optical and electrical characterization techniques. For the optical analyses, Raman spectroscopy was carried out using a T64000 spectrometer (Jobin Yvon Horiba, France) with a 532 nm excitation wavelength. This was performed to compare the Raman spectra of the fabricated device under conditions with and without induced acoustic strain, as illustrated in **Figure 1**. The electrical characterization involved carrying out current-voltage (I-V) measurements using a Keithley 2450 source meter and a DC probe station, allowing for an evaluation of the device's electrical behavior under the same acoustic strain conditions.

*2.2 Computational Modelling*:

To investigate the interactions between the 2D diopside flakes and their interface with a $SiO_2$ substrate, a structural model of the substrate with a crystallographic orientation (001) was created, with approximate dimensions of 115 × 126 $Å^2$. The cleaving process exposes a surface terminated with oxygen atoms covalently bonded to passivate the dangling bonds, resulting in a total thickness of approximately 7.5 Å. The opposite surface, which does not directly interact with the 2D diopside flakes, was passivated with hydrogen atoms. The 2D diopside flakes were generated by cleaving the bulk diopside [15] structure along the (001) plane, yielding a slab with a thickness of approximately 5.2 Å. After passivating the unsaturated bonds, a flake with lateral dimensions of approximately 50 × 20 $Å^2$ was selected. The substrate and diopside flakes were optimized using the Forcite module in the Materials Studio software package [17]. The substrate optimization was performed under periodic boundary conditions along the in-plane directions (*x* and *y*). At the same time, a vacuum layer of 100 Å was introduced along the *z*-direction to avoid spurious interactions between periodic mirror images. The optimization procedure employed the Smart algorithm [17] with convergence criteria set to $2 \times 10^{-5}$ kcal/mol for total energy, 0.001 kcal/mol/Å for maximum atomic force, and 0.001 GPa for residual stress. The simulation cell was fully relaxed under zero external pressure. The Universal force field [18] was employed, with electrostatic interactions treated using the Ewald summation method (accuracy of $1 \times 10^{-5}$ kcal/mol and buffer width of 0.5 Å). The van der Waals interactions were modelled using an atom-based approach with cubic spline truncation, a cutoff distance of 18.5 Å, a spline width of 1.0 Å, and a buffer of 0.5 Å, including long-range correction terms. The (MD) simulations were carried out under an NVT ensemble for a total duration of 50 ps. Initial atomic velocities were randomly assigned to correspond to a target temperature of 298 K. A time step of 1 fs was employed, resulting in 50,000 integration steps of Newton's equations of motion. Temperature control was maintained using a Nosé thermostat [19] with a Q ratio 0.01. During the simulations, the substrate atoms were kept fixed to isolate the thermal response of the 2D flakes. After structural optimization, the $SiO_2$ substrate exhibits Si–O bond lengths of approximately 1.87 Å and O–O surface distances of around 1.38 Å. In the case of the 2D diopside flakes, magnesium forms bonds with silicon (~2.9 Å), oxygen (~2.3 Å), and other magnesium atoms (~2.8 Å). Additionally, the Si–O bonds within the 2D diopside structure measure approximately 1.7 Å. Following MD simulations at room temperature, the equilibrium interlayer distance between the 2D diopside and the substrate stabilizes at approximately 2.26 Å after 50 ps.

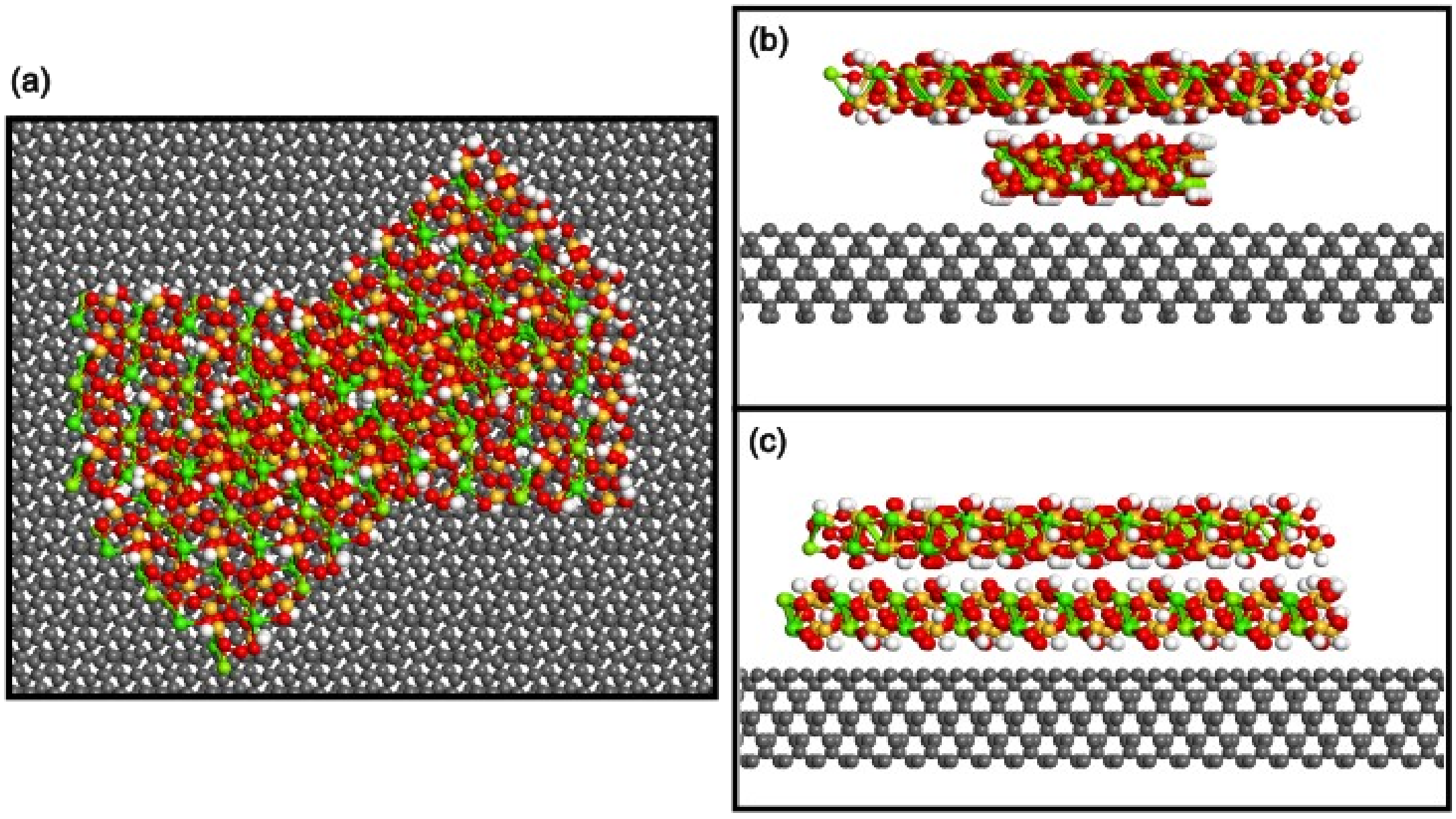

**Figure 2**. Optimized atomic configurations of the $SiO_2$ substrate with two 2D diopside flakes. (a) Frontal view (XY plane) showing the overall arrangement. (b, c) Side views (XZ and YZ planes) highlight the relative rotation between the flakes. Atomic color scale: gray for the substrate (Si and O), white for H, red for O, green for Mg, and orange for Si.

Following the separate optimizations, two identical 2D diopside flakes were positioned above the substrate, with their centers of mass aligned with that of the substrate. The initial interlayer distances between each flake and the substrate and between the flakes were set to 2.5 Å. To evaluate the effect of relative orientation under ambient conditions, we rotated one of the flakes by 45° over the other, as illustrated in **Figure 2**.

3. **Results and Discussion**:

X-ray photoelectron spectroscopy (XPS) was first employed to confirm the elemental composition and chemical states of the surface [20,21], revealing the presence of Ca, Al, Mg, Si, and O atoms. The O 1s spectrum exhibited a dominant oxide contribution at 532 eV, with deconvolution showing that 19.31% of the oxygen species existed as chemisorbed states, while the majority participated in metal-oxide bonds (**Figure 1(c)**). Similarly, the Ca 2p peak at 348 eV indicated $Ca^{2+}$ in the $2p_3/_2$ state, primarily bonded to OH, F, or Cl atoms (**Figure 1(d)**), alongside a smaller Ca $2p_1/_2$ peak at 357 eV and a satellite feature at 336 eV. Meanwhile, Si 2p (103 eV) spectra (**Figure 1(e)**) demonstrated nearly equal contributions from oxide and silicate environments, corroborating the expected chemical structure.

With the 2D nature of d-silicates thus verified, the alignment behavior of 2D diopside flakes under AC electric fields was systematically investigated through combined experimental measurements and computational modeling. **Figure 3(a)** shows the optical image of the 2D diopside flakes deposited on the IDE structure. The SEM image in **Figure 3 (b)** reveals that the flakes are stacked on each other and bridge the gap between the electrodes, forming a continuous active region where the electric field is applied. Each flake exhibits a distinct orientation, which may influence the material's response under the electric field. An alternating current (AC) electric field is applied to these electrodes to induce acoustic strain within the diopside material. The bridging of the 2D diopside flakes ensures a continuous pathway for charge transport and interaction with the applied electric field. The AC electric field generates dynamic vibrations within the material, leveraging its flexoelectric properties.

A Raman spectroscopy study was performed to investigate the vibrational states of the 2D diopsides in the absence and presence of an applied electric field to determine the influence of the field on the material (**Figure 3(c)**). Raman peaks in 2D materials can reveal the presence of strain through shifts in peak positions, changes in intensity, and variations in peak width [22]. For the 2D d-silicate studied here, the peak at 1030 $cm^{-1}$ corresponds to significant stretching of the Ca–O bond. A pronounced Si–O (–Ca) bond vibration was noted at 693.2 $cm^{-1}$. Bending vibrations involving the $[MgO_4]^{6-}$ tetrahedron and Si–O (–Ca) are seen at 420.17 $cm^{-1}$, while the peak at 368 $cm^{-1}$ represents the Si–O bending. The vibration of Si–O (–Mg) bonds was observed at 353 $cm^{-1}$. Peaks at lower frequencies, 302 $cm^{-1}$ and 213 $cm^{-1}$, correspond to symmetric stretching of Ca–O and Si–O(–Ca) bonds, respectively. A significant decrease in peak intensities for all the bonds was observed under the application of an AC electric field with a 10 MHz frequency. This decrease in intensities is assumed to arise due to the acoustic strain induced in the material by the electric field, originating from its flexoelectric [15] properties. Flexoelectricity allows the material to exhibit strain under an applied electric field [23,24], causing vibrational modes within the 2D diopside structure to change their axes. In its initial state, the 2D flakes were randomly oriented, resulting in isotropic scattering axes for Raman measurements. However, the application of the electric field is expected to induce small vibrations in the flakes, leading to their partial alignment. This alignment changes the axes of scattering for Raman signals, reducing the intensity of Raman peaks. The findings support the idea that the electric field modifies the interaction of vibrational modes within the material, which may also contribute to the observed decrease in intensity, particularly at lower frequencies. The Raman spectra analyses were also performed, as shown in **Figure 3(d)**, varying frequencies of the applied AC electric field to investigate its influence on the 2D diopside flakes. A higher decrease in Raman peak intensities was observed for frequencies as the range of frequencies increased from 100 kHz to 10 MHz.

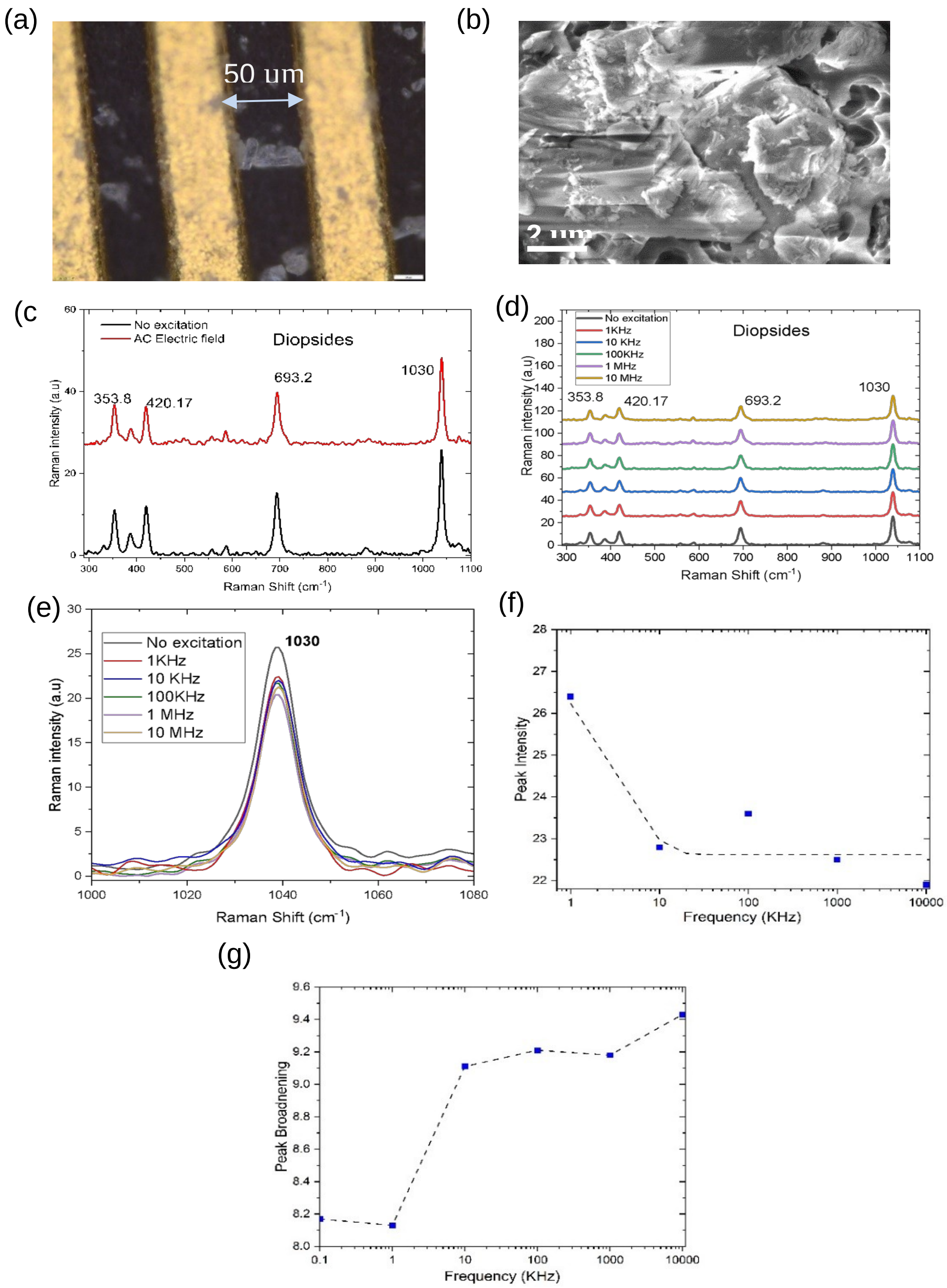

**Figure 3**. (a) Optical image of the 2D diopsides flakes deposited on IDE. (b) SEM image of the 2D diopsides flakes. (c) Raman spectra of 2D diopsides obtained in the absence and presence of an applied AC electric field. (d) Raman spectra of 2D diopsides, illustrating changes with varying AC electric field frequency. (e) A magnified plot of one of the main peaks (1030 $cm^{-1}$) of Raman spectra for 2D diopsides obtained in (d). (f) Variation of Raman peak intensities with respect to the frequency of the AC electric field. (g) Variation of Raman peak broadening with respect to the frequency of the AC electric field.

This behavior can be attributed to the interplay between field-induced vibrations and the response time of the flakes. The change in peak intensities, quantified as a ratio of 1.14-1.25, is evident from the magnified plot of the Raman peak at 1030 $cm^{-1}$ presented in **Figure 3(e)**. The vibrations appear to be sufficiently sustained to facilitate the reorientation of the flakes in these magnified peaks of the Raman spectra, and the rapid oscillations likely outpace the flakes' ability to respond dynamically, resulting in a reduced impact on their orientation and scattering behavior. The magnified plots of the other Raman peaks at 353-693 $cm^{-1}$ were added in the **supplementary information**. A Gaussian fit was applied to the peak at 1030 $cm^{-1}$ (shown in **Figure S4** of the supplementary information) to analyse its intensity and broadening as a function of the applied frequencies. The peak intensity exhibits an exponential decrease with increasing frequency, as shown in **Figure 3(f)**, although the decrease is relatively small. In contrast, the peak broadening initially increases and then stabilizes, as observed in **Figure 3(g)**. This shows an effect in the diopside orientation due to the acoustic strain produced in the flakes by the applied AC electric field.

To further confirm the alignment of the 2D flakes under acoustic strain, current-voltage (I-V) measurements were performed on the IDE structure in the absence and presence of an AC electric field, as shown in **Figure 4(a)**. The I-V characteristics, presented in **Figure 4(b)**, demonstrate a significant increase in current under the strained condition caused by the application of an AC electric field. This increase reflects the reorganization and alignment of the 2D flakes caused by the acoustic strain, which improves charge transport pathways, reduces resistance, and enhances conductivity. These results corroborate the Raman findings and emphasize the role of electric field-induced strain in influencing both the vibrational and electrical properties of the 2D diopside flakes.

For the I-V analyses, however, the increase in current with varying frequency was not as pronounced as shown in the zoomed area of **Figure 4(d)**. This suggests that the reorientation of flakes at these frequencies has a similar effect on the formation of low-resistance pathways for electrical conduction.

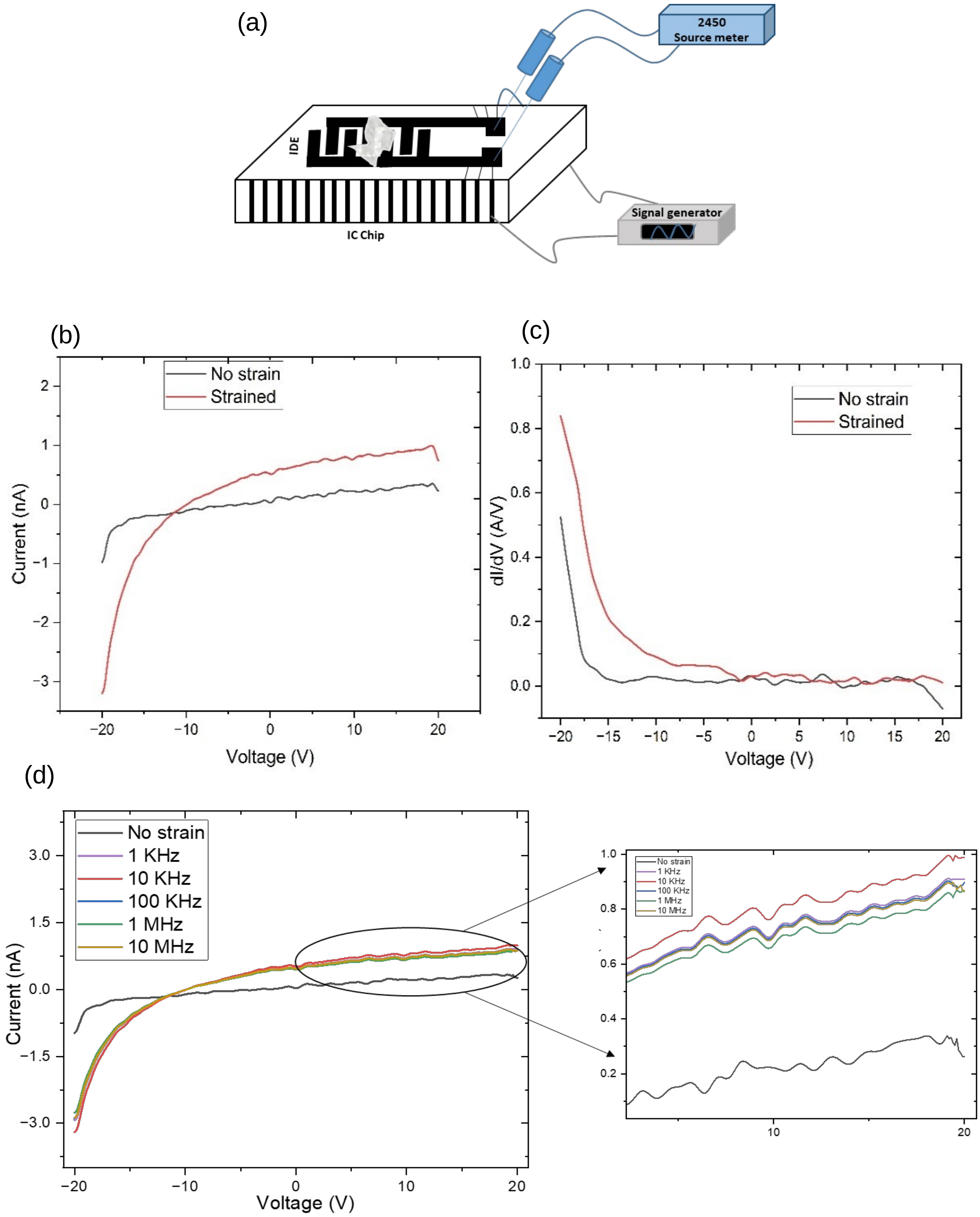

**Figure 4**. (a) DC measuring set up for I-V measurement of the 2D diopside flakes on IDE in the absence and presence of acoustic strain caused by the AC electric field. (b) I-V characteristics of 2D diopside flakes in the absence and presence of strain caused by the applied AC electric field. (c) Conductance of diopsides in the presence and absence of acoustic strain. (d) I-V characterisation of 2D diopsides illustrating changes with varying strain.

These findings imply that while the vibrational alignment at mid-range frequencies is strong enough to impact Raman intensities, it may not be sufficient to modify the pathways for charge transport to produce a significant increase in current. **Figure 4(c)** shows the conductance curve of 2D diopsides in the presence and absence of acoustic strain. The conductance curve shows that the strained 2D diopsides exhibit higher dI/dV values at high negative voltages (-20V to -10V) compared to the unstrained samples, indicating enhanced charge transport due to strain-induced alignment of conduction pathways or reduced carrier scattering. In the moderate voltage range (-10V to 0V), both curves converge to lower conductance, but the strained samples maintain slightly higher values, suggesting partial retention of improved mobility. At positive voltages (0V to 20V), the conductance remains minimal for both cases, implying asymmetric charge transport, likely influenced by interface effects or dielectric constraints. The overall increase in conductance under strain suggests modifications in dielectric properties, leading to enhanced polarizability and possible changes in capacitance. Strain-induced structural realignment in 2D diopsides likely alters electronic states and conduction pathways, facilitating better charge transport.

As mentioned above, we have used fully atomistic MD simulations to gain further insights into the diopside alignments. Our MD results show that there is a progressive alignment of the flakes during the simulation process, driven by an energy exchange within the system. Supplementary information includes MD simulation videos demonstrating the alignment behavior. **Figure 5(a)** shows a representative MD snapshot taken at 5 ps, where the central region of the flake closest to the substrate begins its adsorption process. By 10 ps, as shown in **Figure 5(b)**, this first flake has achieved a nearly planar configuration relative to the substrate. At 20 ps (**Figure 5(c)**), both flakes tend to align, with a clear configuration emerging by 30 ps (**Figure 5(d)**), where the flakes adopt a preferential orientation relative to the substrate.

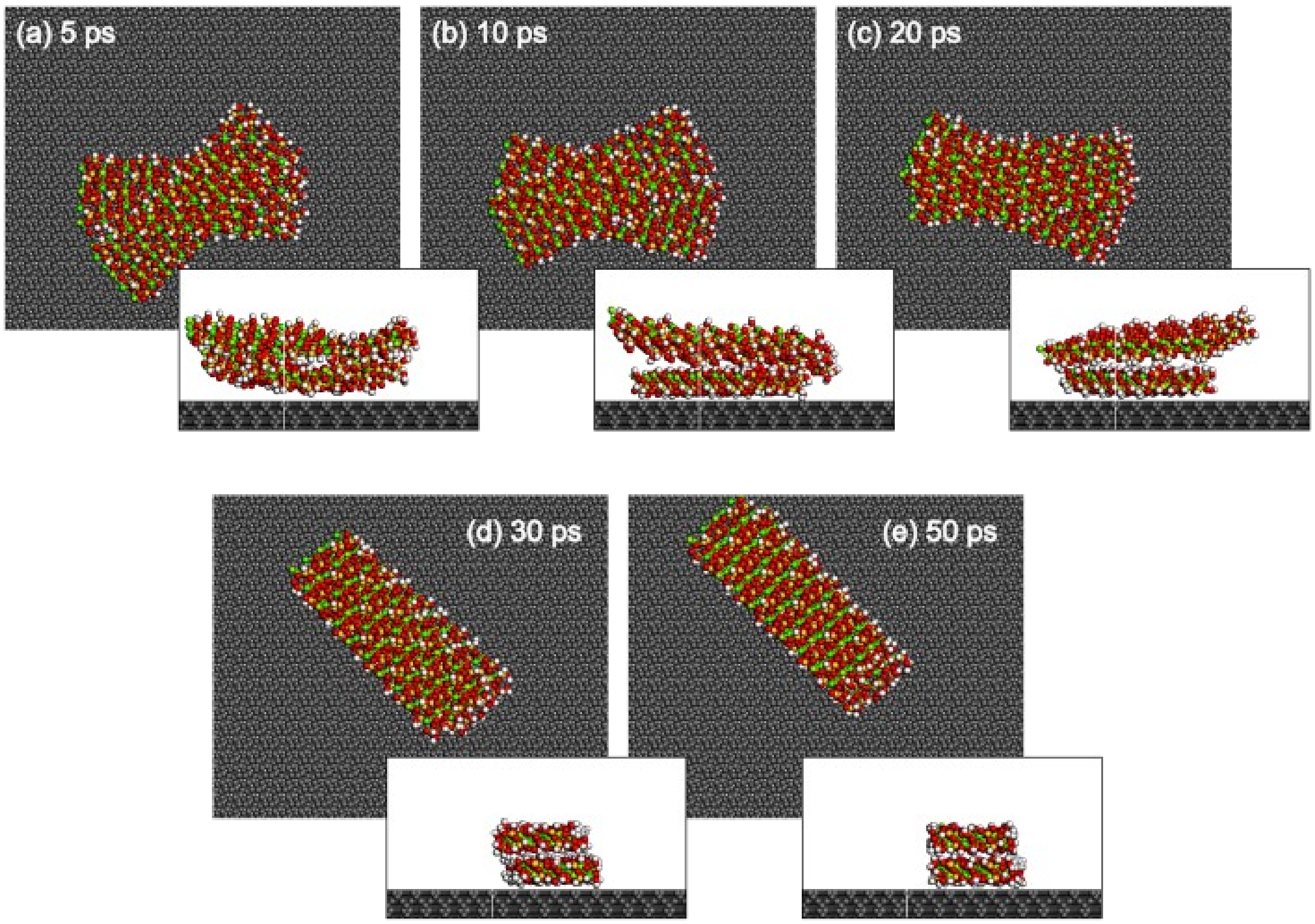

**Figure 5**. Representative MD snapshots taken during the simulation time evolution of the 2D diopside flakes on the $SiO_2$ substrate. (a–e) Snapshots at 5, 10, 20, 30, and 50 ps show the flakes' progressive adsorption, alignment, and stabilization. Inset panels provide side views highlighting structural adjustments over time.

The primary diopside flake aligns perpendicular to these undulations, while the second flake follows the same orientation, stabilizing by the end of the 50 ps simulation (**Figure 5(e)**). The inset panels illustrate the alignment direction of the flakes within the substrate's surface topography. The alignments are energetically favored as the thermal energy allows the flakes to realign/rotate to maximize the van der Waals interactions, increasing structural stability and contributing to increasing the conductivity. A better visualization of the process dynamics can be obtained from the videos in the supplementary information. The process is lock-like, as there is an opposing energy barrier to unlock/unrotate [25]. Thus, there is a continuous alignment/rotation to minimize the structural energy. These findings align well with the experimental results, where the applied AC electric field induced acoustic strain in the 2D diopside flakes, promoting partial alignment. This reorientation alters the scattering axes for Raman signals, leading to a decrease in Raman peak intensities.

## 4. **Conclusions**:

This study demonstrates that AC electric fields induce acoustic strain in 2D diopside flakes via flexoelectric effects, leading to measurable structural alignment/rotation and property modulation. Raman spectroscopy revealed a frequency-dependent reduction in peak intensities, corroborated by enhanced conductivity in I-V measurements, suggesting improved charge transport due to strain-induced flake reorientation. Molecular dynamics simulations further validated alignment dynamics, showing that diopside flakes stabilize in a preferred orientation within 50 ps, with interlayer spacing converging to ~2.12 Å. The combined experimental and computational results underscore the interplay between flexoelectric strain and material response, providing a mechanistic understanding of alignment in 2D silicates. These findings highlight the potential of field-driven strain engineering for tailoring the electronic and vibrational properties of diopside-based systems. Future work will explore higher-frequency regimes, alternative substrates, and heterostructure designs to optimize alignment for applications in flexible electronics, sensors, and energy storage devices.

## 5. Acknowledgements

Guilherme S. L. Fabris thanks the postdoc scholarship financed by the São Paulo Research Foundation (FAPESP) (process number $\#$2024/03413-9) and R.B.O. thanks CNPq process numbers 151043/2024-8 and 200257/2025-0.

B.I. thank CNPq process number \#153733/2024-1, B.I. also thanks FAPESP process number 2024/11016-0 .

M.L.P.J. acknowledges financial support from FAPDF (grant 00193-00001807/2023-16), CNPq (grant 444921/2024-9), and CAPES (grant 88887.005164/2024-00).

Douglas S. Galvão acknowledges the Center for Computing in Engineering and Sciences at Unicamp for financial support through the FAPESP/CEPID Grant (process number $\#$2013/08293-7). We thank the Coaraci Supercomputer for computer time (process number $\#$2019/17874-0) and the Center for Computing in Engineering and Sciences at Unicamp (process number $\#$2013/08293-4).

## 6. References:

[1] Choi S-H 2017 Unique properties of graphene quantum dots and their applications in photonic/electronic devices *J. Phys. D: Appl. Phys.* 50 103002

[2] Tan C, Tang J, Gao X, Xue C and Peng H 2025 2D bismuth oxyselenide semiconductor for future electronics *Nat. Rev. Electr. Eng.*

[3] Kim J H, Jeong J H, Kim N, Joshi R and Lee G-H 2018 Mechanical properties of two-dimensional materials and their applications *J. Phys. D: Appl. Phys.* 52 083001

[4] Lin Z, Carvalho B R, Kahn E, Lv R, Rao R, Terrones H, Pimenta M A and Terrones M 2016 Defect engineering of two-dimensional transition metal dichalcogenides *2D Mater.* 3 022002

[5] Takahashi N and Kuroda K 2011 Materials design of layered silicates through covalent modification of interlayer surfaces *J. Mater. Chem.* 21 14336

[6] Sakuma H and Suehara S 2015 Interlayer bonding energy of layered minerals: Implication for the relationship with friction coefficient *J. Geophys. Res. Solid Earth* 120 2212–19

[7] Xia X, Yih J, D'Souza N A and Hu Z 2003 Swelling and mechanical behavior of poly(N-isopropylacrylamide)/Na-montmorillonite layered silicates composite gels *Polymer* 44 3389–93

[8] Geim A K 2009 Graphene: Status and Prospects *Science* 324 1530–34

[9] Wang F et al 2016 Two-Dimensional Non-Layered Materials: Synthesis, properties and applications *Adv. Funct. Mater.* 27

[10] Novoselov K S and Neto A H C 2012 Two-dimensional crystals-based heterostructures: materials with tailored properties *Phys. Scr.* T146 014006

[11] Sun Y and Liu K 2018 Strain engineering in functional 2-dimensional materials *J. Appl. Phys.* 125

[12] Mahapatra P L, Costin G, Galvao D S, Lahiri B, Glavin N, Roy A K, Ajayan P M and Tiwary C S 2023 A comprehensive review of atomically thin silicates and their applications *2D Mater.* 11 032003

[13] Lin F et al 2018 Planar alignment of graphene sheets by a rotating magnetic field for full exploitation of graphene as a 2D material *Adv. Funct. Mater.* 28

[14] Lim W P, Yao K and Chen Y 2007 Alignment of carbon nanotubes by acoustic manipulation in a fluidic medium *J. Phys. Chem. C* 111 16802–07

[15] Mahapatra P L et al 2023 Energy harvesting using two-dimensional (2D) d-silicates from abundant natural minerals *J. Mater. Chem. C* 11 2098–2106

[16] Mahapatra P L et al 2024 3D-printed flexible energy harvesting devices designed using non-layered two-dimensional natural tourmaline silicates *J. Mater. Chem. C* 12 3418–29

[17] Meunier M and Robertson S 2021 Materials Studio 20th anniversary *Mol. Simul.* 47 537–39

[18] Rappe A K, Casewit C J, Colwell K S, Goddard W A and Skiff W M 1992 UFF, a full periodic table force field for molecular mechanics and molecular dynamics simulations *J. Am. Chem. Soc.* 114 10024–35

[19] Nosé S 1984 A molecular dynamics method for simulations in the canonical ensemble *Mol. Phys.* 52 255–68

[20] Hochella M F and Brown G E 1988 Aspects of silicate surface and bulk structure analysis using X-ray photoelectron spectroscopy (XPS) *Geochim. Cosmochim. Acta* 52 1641–48

[21] Eggleston C M, Hochella M F and George P A 1989 Sample preparation and aging effects on the dissolution rate and surface composition of diopside *Geochim. Cosmochim. Acta* 53 797–804

[22] Pimenta M A, Del Corro E, Carvalho B R, Fantini C and Malard L M 2014 Comparative study of Raman spectroscopy in graphene and MOS2-type transition metal dichalcogenides *Acc. Chem. Res.* 48 41–47

[23] Jia X, Guo R, Chen J and Yan X 2024 Flexoelectric effect in thin films: Theory and applications *Adv. Funct. Mater.*

[24] Cross L E 2006 Flexoelectric effects: Charge separation in insulating solids subjected to elastic strain gradients *J. Mater. Sci.* 41 53–63

[25] Otero R et al 2004 Lock-and-key effect in the surface diffusion of large organic molecules probed by STM *Nat. Mater.* 3 779–82